\pdfoutput=1
\documentclass[
  draft
]{agujournal2025}

\usepackage{url} 
\usepackage{lineno}
\usepackage{soul}
\let\linenumbers\relax

\usepackage{amsmath}
\usepackage{amssymb}
\usepackage{mathtools}
\usepackage{physics}
\usepackage{graphicx}
\usepackage{bm}
\usepackage{pifont}
\usepackage{wasysym}
\usepackage{xcolor}
\usepackage{contour}
\contourlength{0.5pt} 
\contournumber{10} 

\draftfalse

\begin{document}

\journalname{JGR: Planets}

\title{Scaling of impact ripples across planetary conditions}

\authors{C. W. Lester\affil{1,2}, O. Duran\affil{3}, A. B. Murray\affil{1}}

\affiliation{1}{Earth and Climate Sciences, Duke University, Durham, NC, USA}
\affiliation{2}{Physics and Physical Oceanography, University of North Carolina Wilmington, Wilmington, NC, USA}
\affiliation{3}{Department of Ocean Engineering, Texas A\&M University, College Station, TX, USA}

\correspondingauthor{Conner W. Lester}{conner.lester@duke.edu, lesterc@uncw.edu}

\authoraddr{%
                      Conner W. Lester (conner.lester@duke.edu, lesterc@uncw.edu)
                   }

\authorrunninghead{Ignored.} 

\keypoints%
    {Simulated impact ripple wavelengths are set by the grain diameter for sand transport close to threshold regardless of the fluid density.}
    {Simulated impact ripple wavelengths increase with the fluid shear velocity for intense transport conditions.}
    {Impact ripples migrate against the wind as antiripples for low enough density ratios, such as on Venus, or large enough grains on Earth.}

\titlerunninghead{Ignored.}

\maketitle

\begin{abstract}

Ubiquitous wind-driven (aeolian) ripples have been thought to emerge solely from grain collisions with the bed during saltation. Recent findings, however, have shown that there are two scales of aeolian ripples, with the larger scale having a hydrodynamic origin, akin to water ripples, and the smaller scale (`impact ripples') being hypothesized to emerge from the grain-bed collision mechanism. Still, there are several instances where only one ripple scale is found and the formation mechanism is not clear. Here we compare numerical simulations of only impact ripples with observations, over a broad range of planetary conditions, to clarify their defining characteristics. We find two main regimes for the scaling of impact ripple wavelengths based on transport intensity: for low-intensity transport (`saltation' regime), impact ripple wavelengths scale with grain size regardless of the fluid density; whereas for high-intensity transport (`collisional' regime) wavelengths scale with the average hoplength and thus increase with the fluid velocity. In contrast, impact ripple migration speeds strongly depend on the grain-fluid density ratio, grain size and fluid velocity. Finally, we find that impact ripple speeds become zero or negative (denoting stationary ripples and antiripples, respectively) for grain-fluid density ratios below a threshold that depends on the grain size. In particular, for low density ratios, like those found on Venus, or millimeter-scale grains on Earth, we find only antiripples. Our results thus provide a general framework for interpreting ripple emergence across environments and for identifying impact ripples in ambiguous field or experimental data.
     \begin{plainlanguagesummary}
Wind-blown sand forms ripples on Earth, Mars, and other planetary bodies. Recent experiments found two coexisting types of wind ripples: larger ``hydrodynamic ripples'', likely shaped by the fluid flow like underwater ripples, and smaller ``impact ripples'' formed by grains striking the sand bed. Observations are sparse and often report only one ripple size, making their origins uncertain. To isolate impact ripple behavior across planetary environments, we simulated impact ripples from Mars-like to Venus-like conditions, across a range of grain sizes and wind speeds, and compared them with observations. At low wind speeds, impact ripple wavelengths are fixed by the grain size (centimeter scale for fine sand) across planets. In contrast, ripple speeds vary strongly with atmospheric density, grain size, and wind speed. In dense atmospheres, like those of Venus, or for coarse sand on Earth, impact ripples can even move against the wind as ``antiripples'', which have yet to be observed in nature. These results assist our interpretation of wind-blown ripples and guide future observations of both ripple types, including where to look for the first antiripples.
     \end{plainlanguagesummary}
     %
\end{abstract}

\section{Introduction}

Aeolian ripples are a common emergent pattern of wind-driven sediment transport. They are observed in arid environments on Earth and Mars \cite{lapotre2016large} and have been widely studied in wind-tunnel experiments spanning ambient terrestrial conditions \cite{andreotti2006aeolian,cheng2018experimental,schmerler2016experimental,walker1981experimental,alvarez2025ripples,yizhaq2024coevolving}, high Venus-like pressures \cite{greeley1984microdunes}, and low Martian-like pressures \cite{yizhaq2024coevolving,miller1987wind,andreotti2021lower,alvarez2025ripples}. Across these environments, ripple wavelengths $\lambda$ for unimodal grain diameters $d$ range from centimeters to decimeters and even meters in the case of Mars. This variability is generally attributed to differences in atmospheric density, grain size, and wind strength. However, the dependence of ripple wavelengths on these parameters remains unclear, and observations across environments often appear contradictory.

Wind-tunnel experiments for small grains ($d\sim 100\ \mu$m), under conditions relevant to both Earth and Mars, have reported the coexistence of two distinct ripple scales: small ripples with wavelengths of order 1 cm and larger bedforms with wavelengths of order 10 cm or larger \cite{yizhaq2024coevolving,alvarez2025ripples,miller1987wind}. \citeA{yizhaq2024coevolving} and \citeA{alvarez2025ripples} proposed that the smaller bedforms are `impact' ripples, whereas the larger ones are `hydrodynamic' ripples (in what follows we refer to `hydrodynamic' for both water and air flows), similar to those in water \cite{lapotre2016large,duran2019unified}. A similar dynamical distinction has been proposed for Martian bedforms, where small impact ripples coexist with much larger meter-scale bedforms that are also hypothesized to arise from hydrodynamic mechanisms \cite{lapotre2016large,duran2019unified}. These observations suggest that multiple ripple-forming mechanisms operate simultaneously across planetary environments \cite{yizhaq2024coevolving}.

\begin{figure}[t!]
\noindent\includegraphics[width=\textwidth]{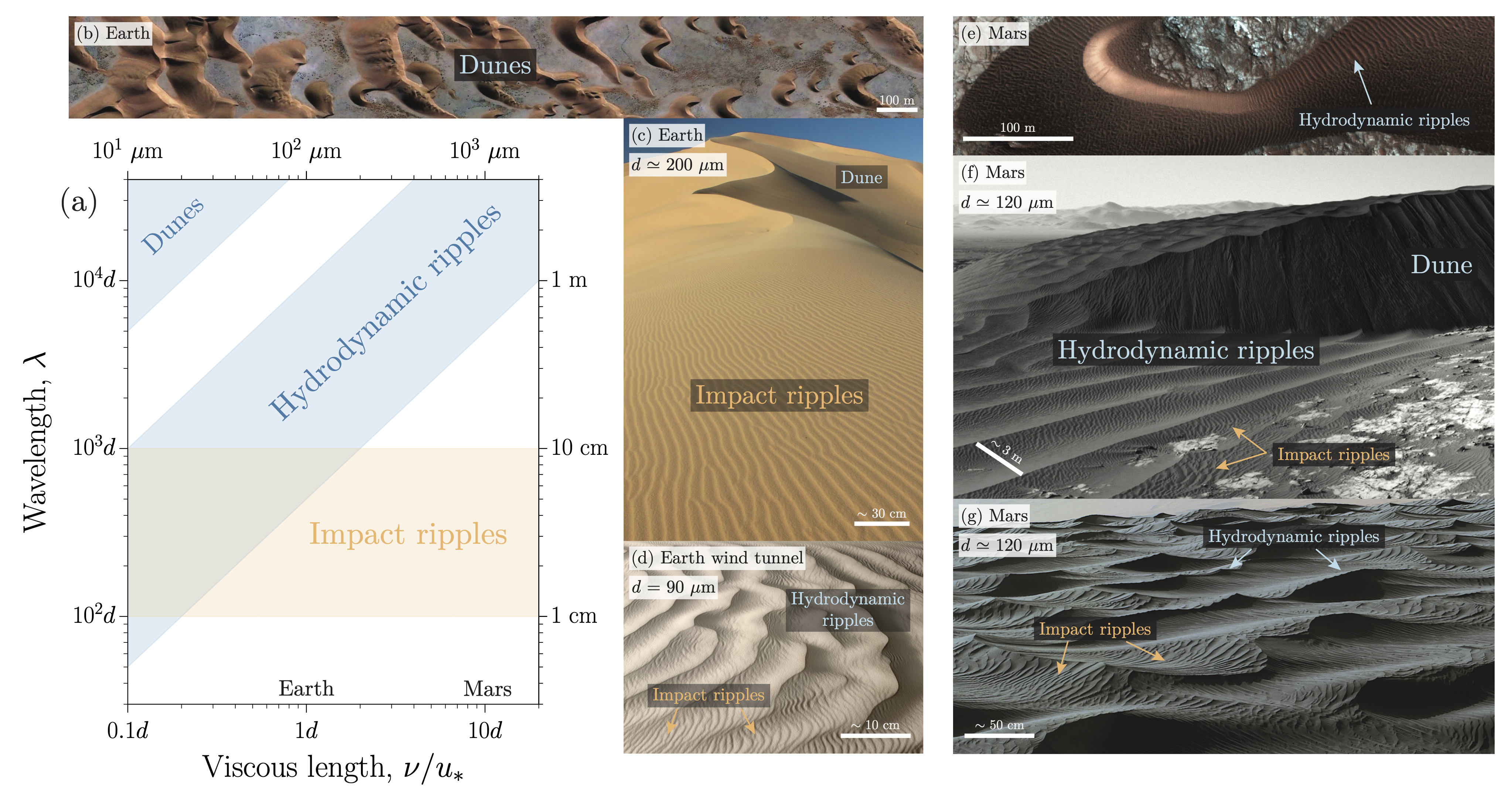}
\caption{Bedform regimes across environments. (a) Typical range of bedform wavelengths $\lambda$ as a function of the viscous length $\nu/u_*$, based on theory and observations \cite{lapotre2016large,duran2019unified,yizhaq2024coevolving}. Hydrodynamic and impact ripple regimes overlap for Earth-like conditions but remain well separated for Mars-like conditions. (b-d) Images of bedforms on Earth. (b) Dunes in West Sahara (Google, Digital Globe). (c) Ripples on a dune in Morocco---thought to be impact ripples \cite{andreotti2006aeolian}. (d) Ambient air wind tunnel experiments for small grains \cite{yizhaq2024coevolving}. (e-g) Images of bedforms on Mars. (e) Martian dune in Nili Patera dune field (HiRISE image). (f,g) Two ripple sizes on Namib Dune, Gale Crater (Mastcam mosaic, Curiosity rover). Note that Earth wind tunnel experiments with small grains (d) and Martian imagery (f,g) both show the coexistence of impact and hydrodynamic ripples \cite{yizhaq2024coevolving,alvarez2025ripples}, while field ripples in Morocco with larger grains (c) show only one scale of ripples. }
\label{fig1}
\end{figure}

Hydrodynamic ripples form through modulations of the fluid flow over a bedform and the resulting lagged response of the sediment flux to this modulation \cite{duran2019unified,charru2013sand}. Under hydrodynamically smooth and transitional conditions, hydrodynamic ripples and dunes can coexist, with wavelengths separated by nearly an order of magnitude because of a bedform gap associated with a so-called hydrodynamic anomaly. Smooth conditions occur when bed grains are embedded within the viscous sublayer, $d\ll \nu/u_*$, where $\nu/u_*$ is the viscous length scale set by the fluid kinematic viscosity $\nu$ and shear velocity $u_*$. In this regime, the range of hydrodynamic bedform wavelengths is set by the viscous length, $\lambda_{\text{hydro}}\propto \nu/u_*$, implying characteristic ripple sizes of order 10 cm on Earth and order 1 m on Mars (Fig.~\ref{fig1}) \cite{yizhaq2024coevolving,alvarez2025ripples,lapotre2016large}. For $d\gg \nu/u_*$, the flow becomes hydrodynamically rough and only dunes are expected \cite{duran2019unified}.

Although impact ripple formation is not as well understood as hydrodynamic bedforms, it is known that impact ripples form through the basic mechanics of aeolian sediment transport: wind drag accelerates grains in transport that cause ejections of bed-bound grains upon impact. The flat-bed instability behind ripple formation then arises from the coupled impact-ejection mechanics and trajectory modulations over small bed perturbations. 

There is a long history of modeling efforts to understand how the dynamics of sediment transport give rise to impact ripples. Most notably, \citeA{anderson1987theoretical} adopted insights from \citeA{Bagnold1941}, \citeA{ungar1987steady} and others to set the stage for understanding ripple emergence by simplifying sediment transport into two populations: long trajectory rebounding grains, `saltons', and their short trajectory ejecta, `reptons'. This salton-repton model has been refined and expanded on in more recent impact ripple models \cite{prigozhin1999nonlinear,csahok2000dynamics}. The common prediction from such models is that impact ripple wavelengths scale with the repton hoplength, as previously suggested by \citeA{Bagnold1941}. Later, \citeA{duran2014direct} used a model resolving grain-grain and grain-fluid interactions~\cite{duran2012numerical} to simulate ripple formation at relatively high wind speeds compared to the threshold, finding that ripple wavelengths scale with the hoplength averaged over all particle trajectories, not only reptons'. However, because the average hoplength increases as the atmospheric density $\rho_f$ decreases~\cite{duran2014direct}, the prediction is inconsistent with observations that impact ripples have similar sizes as the fluid density varies \cite{yizhaq2024coevolving,alvarez2025ripples,miller1987wind} (Fig.~\ref{fig1}). 
\citeA{lester2025emergence} revised this problem by showing that sediment transport is not naturally separated into two discrete salton and repton populations. Instead, both discrete element method (DEM) simulations \cite{lester2025emergence,duran2014direct} and wind-tunnel experiments \cite{ho2014aeolian,rasmussen2015laboratory} indicate a continuous, approximately scale-free hoplength distribution. In that case, there is no well-defined repton hoplength to scale the ripple wavelengths with. The proposed analytical theory identified a different wavelength selection mechanism with a new associated length scale: the average separation between where an energetic grain impacts the bed and where the ejection of the splashed bed grains occurs. This impact-ejection lag was found to be of the order of the grain diameter, leading to the prediction $\lambda\propto d$ for low-intensity transport conditions close to the threshold. This seemed to contradict previous predictions that impact ripple wavelengths are selected by a transport length scale (the average hoplength~\cite{duran2014direct}) and suggested wavelengths should be similar across planetary environments, at least close to the transport threshold, in agreement with low-pressure wind tunnel observations \cite{yizhaq2024coevolving,alvarez2025ripples,miller1987wind}. \citeA{lester2025emergence} also predicted that the ripple speed $c$ depends sensitively on the planetary environment, with impact ripples migrating against the wind as `antiripples' in dense atmospheres such as that of Venus---although this phenomenon has not yet been observed outside of impact ripple simulations.

The expected regimes of bedforms in unidirectional flows, and infinite flow depth, are summarized in Fig.~\ref{fig1}. On Mars, the impact and hydrodynamic ripple regimes are expected to be well separated, with meter-scale hydrodynamic ripples much larger than centimeter-scale impact ripples (Fig.~\ref{fig1}a,f,g). On Earth, by contrast, the hydrodynamic and impact-ripple regimes can overlap, allowing both bedforms to coexist at similar scales and potentially interact (Fig.~\ref{fig1}a,d). This overlap complicates the interpretation of aeolian ripple observations, especially for small grains, and may explain the large variability of reported ripple wavelengths. To help disentangle these observations, we combine grain-scale impact ripple simulations, analytical predictions, and a wide range of observational data to provide a coherent picture of how impact ripple sizes and migration rates scale across environmental conditions, including where antiripples may occur. It is important to note that the coupled grain-fluid model we use~\cite{duran2012numerical} cannot produce hydrodynamic ripples because it does not resolve the bed shear stress modulation that drive the morphodynamic instability~\cite{duran2019unified}. This makes it an ideal tool to study only impact ripple characteristics, without the ambiguity of experimental and field observations. 

\section{Simulations of sand transport and formation of impact ripples}\label{sec: IR sims}

Sediment transport, and the subsequent formation of impact ripples, is simulated using a discrete element method (DEM) model for grain-grain collisions coupled with a simplified Reynolds-averaged Navier-Stokes equation for the fluid boundary layer driving transport \cite{duran2012numerical}. Despite its simplified fluid dynamics, the coupled model quantitatively reproduces fundamental sediment transport characteristics such as the threshold shear velocity $u_\text{th}$ for sustained transport \cite{pahtz2023scaling}, the mass transport rate $Q_{m}$ [M/LT], the vertical mass flux at the bed $\varphi_{m}$ [M/L$^2$T], probability distributions such as for the hoplength $P(\ell)$, and average trajectory-based quantities such as the average hoplength $\langle\ell\rangle=\int\ell P(\ell)\dd\ell=Q_{m}/\varphi_{m}$ \cite{pahtz2020unification,pahtz2023scaling,lester2025emergence}. Additionally, the wavelengths of the simulated impact ripples were found to agree with low-pressure wind tunnel experiments and Martian field observations \cite{yizhaq2024coevolving,miller1987wind}---suggesting that the model captures the fundamental processes behind the formation and scaling of impact ripples in nature. 

The relevant model parameters are the grain diameter $d$ and density $\rho_p$, the fluid density $\rho_f$ and dynamic viscosity $\mu=\rho_f\nu$, the fluid shear (or friction) velocity $u_*$ and the gravitational acceleration $g$. These parameters can be organized into three dimensionless numbers, the grain-fluid density ratio, a rescaled grain diameter, and the rescaled bed shear stress or Shields number $\Theta$:
 \begin{equation}
 \label{eq: params}
     \frac{\rho_p}{\rho_f},\quad \frac{d}{d_\mu},\quad \Theta =\frac{\rho_fu_*^2}{\rho_p\tilde{g}d},
 \end{equation}
where $d_\mu =\left(\mu/(\rho_p\sqrt{\tilde{g}})\right)^{2/3}$ is a viscous grain scale and $\tilde{g}=(1-\rho_f/\rho_p)g$ is the buoyancy reduced gravity \cite{pahtz2023scaling}. Because the viscous grain scale $d_\mu$ does not depend on the fluid density, and thus is relatively constant across planetary environments ($d_\mu\sim 2\ \mu$m), it allows a clear separation in the physical interpretation of the different numbers: a given planetary environment is mainly described by the density ratio (along with gravity), the rescaled grain diameter ($d/d_\mu$) uniquely describes the grain size, whereas the Shields number describes the sediment transport intensity.

\subsection{`Saltation' and `collisional' transport regimes} 
\label{sub:saltation_and_collisional_transport_regimes}

Steady-state transport is described by the mass transport rate $Q_{m}$, the vertical mass flux at the bed $\varphi_{m}$, and the average hoplength of grain trajectories $\langle\ell\rangle$. Following \citeA{pahtz2023scaling}, we introduce the rescaled transport rate,
 \begin{equation}
 \label{eq: Q*}
     Q_*\equiv \frac{\phi_b Q}{\sqrt{gd^3}}\left(\frac{\rho_p}{\rho_f}\right)^{-1/3},
 \end{equation}
where $Q=Q_m/(\rho_p\phi_b)$ [L$^2$/T] is the volumetric sand flux at bed packing fraction $\phi_b\sim 0.5$. Dividing by the factor $(\rho_p/\rho_f)^{1/3}$ makes $Q_*$ depend mainly on the dimensionless excess shear stress $\Theta - \Theta_{\text{th}}$ \cite{pahtz2023scaling} as
\begin{equation}
    Q_*(\Theta) = a (\Theta - \Theta_{\text{th}}) \left[ 1 + b (\Theta - \Theta_{\text{th}}) \right]
    \label{eq:q}
\end{equation}
where $a$ and $b$ are empirical constants that can depend weakly on $\rho_p/\rho_f$ and $d/d_\mu$ (\citeA{pahtz2023scaling} report $a\simeq 1.7,\ b\simeq 7$) and $\Theta_{\text{th}}$ is the Shields number at the cessation transport threshold \cite{pahtz2018cessation}.

\citeA{pahtz2020unification} identified two different regimes within Eq.~\ref{eq:q}. Close to the transport threshold, the rescaled transport rate is proportional to the dimensionless excess shear stress ($\Theta - \Theta_{\text{th}}$) as the energy of saltating grains is mostly dissipated during impacts with the static bed. This low-intensity transport regime is consistent with standard saltation transport. In contrast, far enough from the threshold, the relation becomes nonlinear (with an effective scaling exponent around 1.5 and growing asymptotically to 2) as midair grain collisions are increasingly important. This `collisional' high-intensity transport regime also describes the standard bedload transport and is representative of the so-called Bagnold regime~\cite{duran2011aeolian,Bagnold1941}. In what follows, we will use the condition $Q_*\simeq 0.1$ (or $\Theta - \Theta_{\text{th}} \simeq 0.045$) to qualitatively separate the `saltation' ($Q_*\lesssim 0.1$) from the `collisional' ($Q_*\gtrsim 0.1$) regime. 

\begin{figure}[t!]
    \noindent\includegraphics[width=\textwidth]{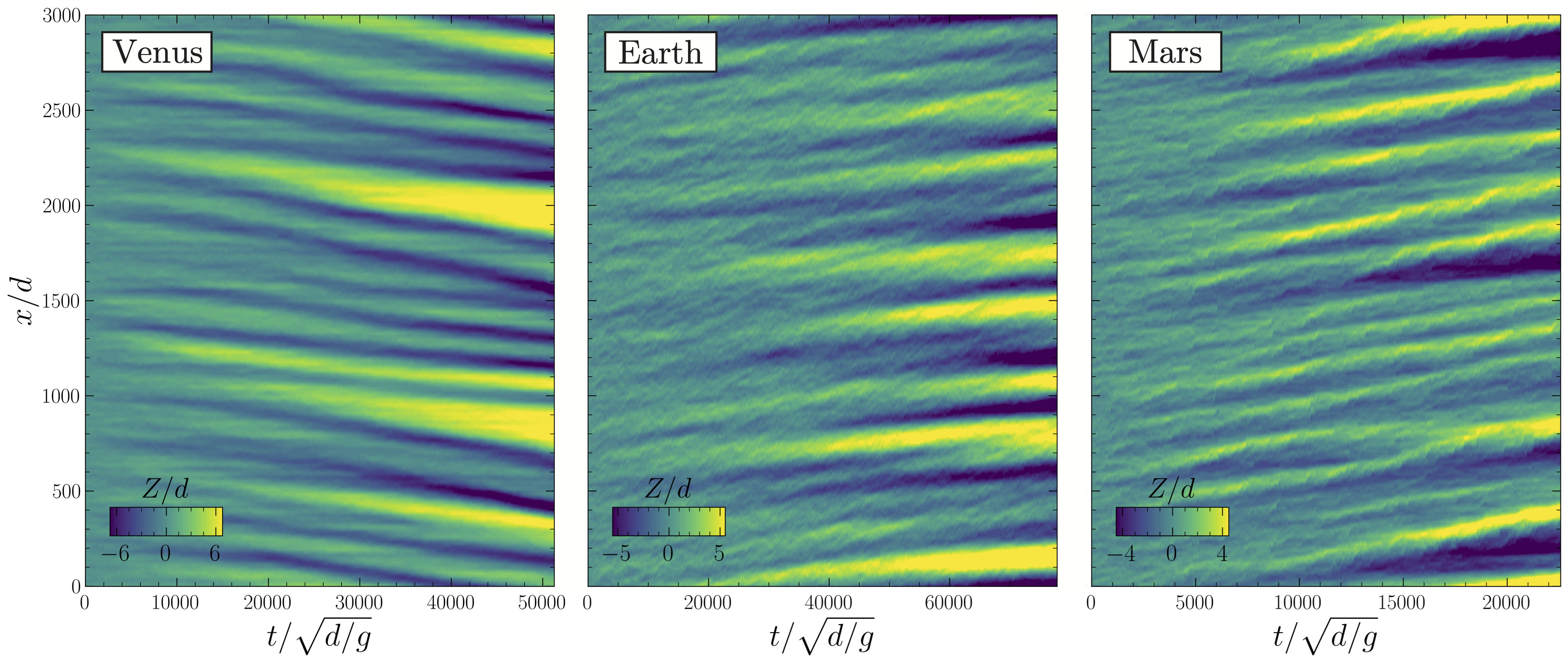}
    \caption{Space-time evolution of DEM simulated impact ripples on Venus-like ($\rho_p/\rho_f=50$), Earth ($\rho_p/\rho_f=2000$), and Mars-like ($\rho_p/\rho_f=10^5$) conditions for grain sizes $d\sim 100\ \mu$m. Wind is moving in the $x$ direction. Space and time are rescaled by the grain size $d$ and gravity $g$. Colors show rescaled surface elevation $Z/d$. Note that Venus ripples are propagating in the opposite direction of the wind as antiripples.}
    \label{fig2}
\end{figure}

\subsection{Simulation of impact ripples}

As previously stated, we use the DEM model to explore impact ripple formation for a wide range of conditions (Eq.~\ref{eq: params}): density ratios $\rho_p/\rho_f$ from water-like ($2.65$) to Mars-like ($2\times 10^5$); rescaled grain diameters $d/d_\mu$ in the range $10$ to $10^3$ (roughly from $20\ \mu$m to $2$ mm); and Shields numbers $\Theta$ corresponding to a rescaled transport rate $Q_*$ in the range $10^{-3}$ to $1$. Simulations begin from a flat bed, and once transport becomes self-sustained, ripples emerge from a linear instability mechanism during which ripple amplitudes grow exponentially over time while their wavelengths $\lambda$ and propagation speeds $c$ remain nearly constant (Fig.~\ref{fig2}) \cite{lester2025emergence,duran2014direct}.

\begin{figure}[t!]
    \noindent\includegraphics[width=\textwidth]{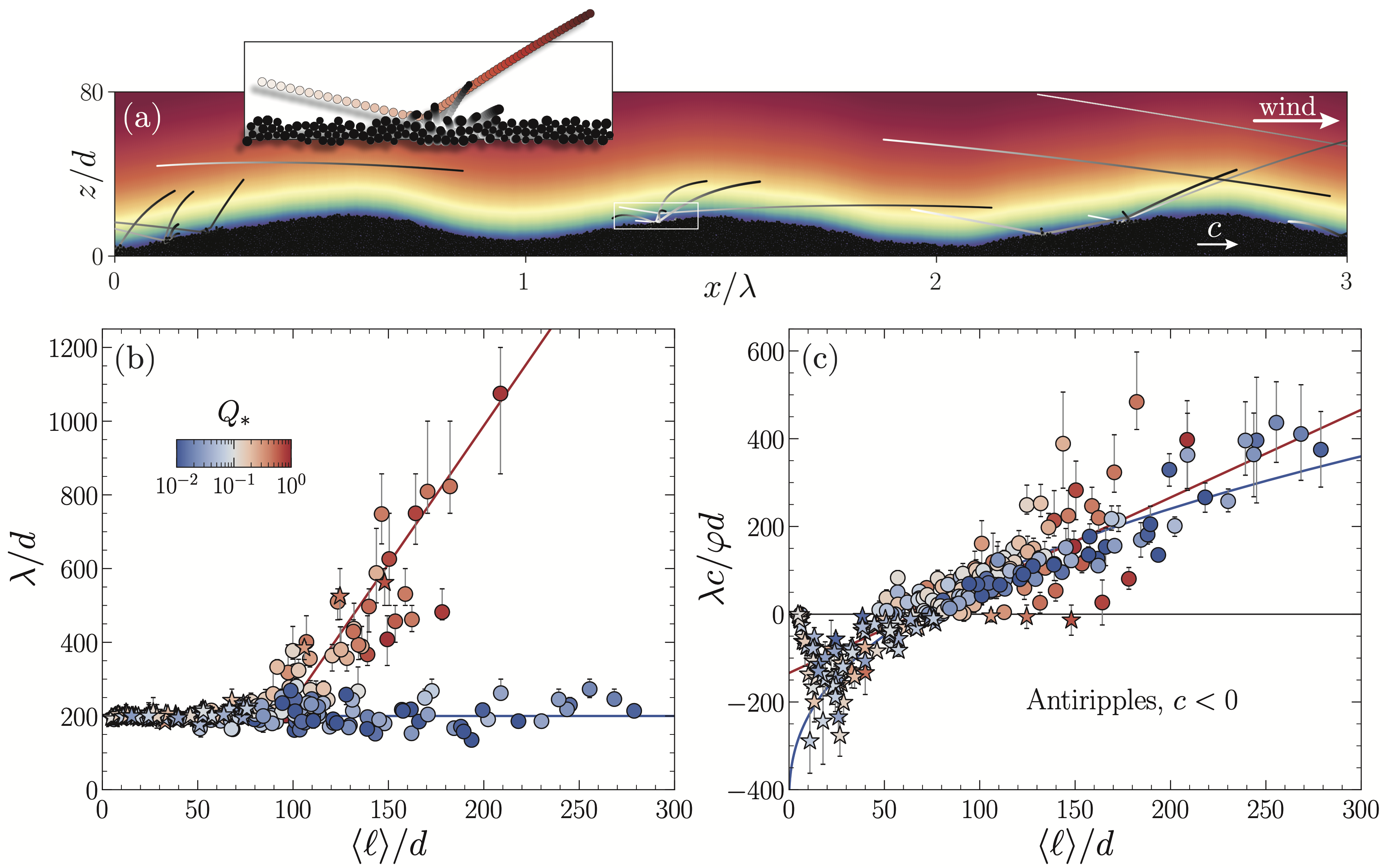}
    \caption{Impact ripple dynamics---numerical simulations and analytical predictions. (a) Snapshot of a DEM simulation of aeolian transport and ripple formation for low-intensity ($Q_*<0.1$) Earth-like conditions ($\rho_p/\rho_f=2000,\ d\simeq 100\ \mu$m, $\lambda\simeq200d$). Grains are shown as black spheres with time-lagged trajectories. Background color shows wind speed increasing from the surface. Inset shows a high energy grain impacting the surface---the main mechanism leading to impact ripple emergence for low-intensity transport $Q_*<0.1$~\cite{lester2025emergence}. (b,c) Simulated impact ripple wavelengths $\lambda/d$ and ripple flux $\lambda c/\varphi d$ versus average hoplength $\langle\ell\rangle/d$ for different transport intensities $Q_*$. Lengths are rescaled by the grain size $d$ and ripple speeds by the vertical flux of the surface $\varphi$. Circles are ripples that propagate with the wind and stars are antiripples ($c<0$). Blue lines are analytical scaling predictions for low-intensity ($Q_*<0.1$) and saltation-like transport ($\rho_p/\rho_f\gtrsim10^2$, or $\langle\ell\rangle\gtrsim 30d$) \cite{lester2025emergence}: $\lambda\propto d$ and $ c/\varphi\propto (\langle\ell\rangle/\langle\ell\rangle_c)^{0.4}-1$, where $\langle\ell\rangle_c\simeq 67d$. Red lines are empirical scaling predictions for intense saltation ($Q_*>0.1$) \cite{duran2014direct}: $\lambda\propto \langle\ell\rangle$ and $c\propto \varphi$, thus $\lambda c\propto \langle\ell\rangle\varphi =Q$.}
    \label{fig3}
\end{figure}

\section{General characteristics of impact ripples}\label{sec: IR theory}

Simulations show that in the `saltation' regime ($Q_*\lesssim0.1$) the wavelength is just proportional to the grain diameter $d$ (Fig.~\ref{fig3}b), a scaling that also holds for all simulations with density ratios in the range $\rho_p/\rho_f < 100$. This scaling originates from the spatial lag between impacting and ejected grains (Fig.~\ref{fig3}a) as predicted by \citeA{lester2025emergence}. The scaling with the grain diameter breaks down in the `collisional' regime ($Q_*\gtrsim0.1$), where the scale of the wavelength is dominated by the average hoplength $\langle\ell\rangle$ (Fig.~\ref{fig3}b). This was found empirically by \citeA{duran2014direct}, whose simulations were performed in what we now call the `collisional' regime. The physical mechanism underlying the wavelength scaling in the `collisional' regime remains poorly understood. However, \citeA{duran2014direct} suggested that the increased collision rate created a gas-like layer near the surface that allowed for gravity-driven diffusion, which was enough to qualitatively explain the scaling. Another possibility is that the gas-like behavior prevents the spatial correlation between local impacts and ejections thus leaving the average hoplength as the only relevant length scale.

The general scaling of ripple propagation speed is more subtle. Because ripple growth and migration rates are mediated by the exchange of grains with the bed, we expect ripple speeds to scale with the vertical flux $\varphi=\varphi_m/(\rho_p\phi_b)$. In the `collisional' (intense) regime, where $\lambda\propto\langle\ell\rangle$, this implies that the product between ripple wavelength and speed is proportional to the transport rate ($\lambda c\propto \langle\ell\rangle\varphi=Q$), as shown by \citeA{duran2014direct}. In the `saltation' (low-intensity) regime, however, \citeA{lester2025emergence} predicted that the rescaled propagation speeds $c/\varphi$ collapse onto an almost unique function of $\langle\ell\rangle/\lambda$, and can become negative for sufficiently short hoplengths. Substituting the scaling $\lambda \propto d$ in this regime, we get 
\begin{equation}
    \label{eq: c}
    \frac{c}{\varphi}\propto \left(\frac{\langle\ell\rangle}{\langle\ell\rangle_c}\right)^\gamma-1,
\end{equation}
where $\gamma\simeq 0.4$ and $\langle\ell\rangle_c\simeq 67d$ is a critical hoplength below which $c<0$ (Fig.~\ref{fig3}c). This prediction of upwind-migrating antiripples is a robust outcome of the analytical theory and simulations (Fig.~\ref{fig2},~\ref{fig3}c) that has yet to be confirmed experimentally or in the field. 

Below we parameterize this ripple speed prediction (Eq.~\ref{eq: c}) in terms of transport parameters (Eq.~\ref{eq: params}) using scaling relations for $\langle \ell\rangle$ and $\varphi$ found in the DEM: the average hoplength for low-intensity saltation ($Q_*\lesssim 0.1, \rho_p/\rho_f\gtrsim 10^2$) scales as $\langle\ell\rangle/d\simeq 40 (\rho_p/\rho_f)^{1/5}(d/d_\mu)^{-0.15}$, and the vertical flux scales as $\varphi/\sqrt{gd}\simeq (\Theta-\Theta_\text{th})/2$. We use the fixed threshold Shields number $\Theta_\text{th}=0.008$ derived from observations \cite{pahtz2018cessation}. Using $Q=\langle\ell\rangle \varphi$, these scaling relations are broadly consistent with low-intensity transport rate scaling in Eq.~\ref{eq:q} \cite{pahtz2023scaling}. However, they give a weak parameter dependence of the prefactor $a\sim (\rho_p/\rho_f)^{-0.13}(d/d_{\mu})^{-0.15}$ because we measure $\langle \ell\rangle$ and $\varphi$ from grains crossing a fixed elevation ($0.7d$ above the bed, as in \citeA{lester2025emergence}), compared to the integral measurement of $Q$ in \citeA{pahtz2023scaling}.

\begin{figure}[t!]
    \noindent\includegraphics[width=.7\textwidth]{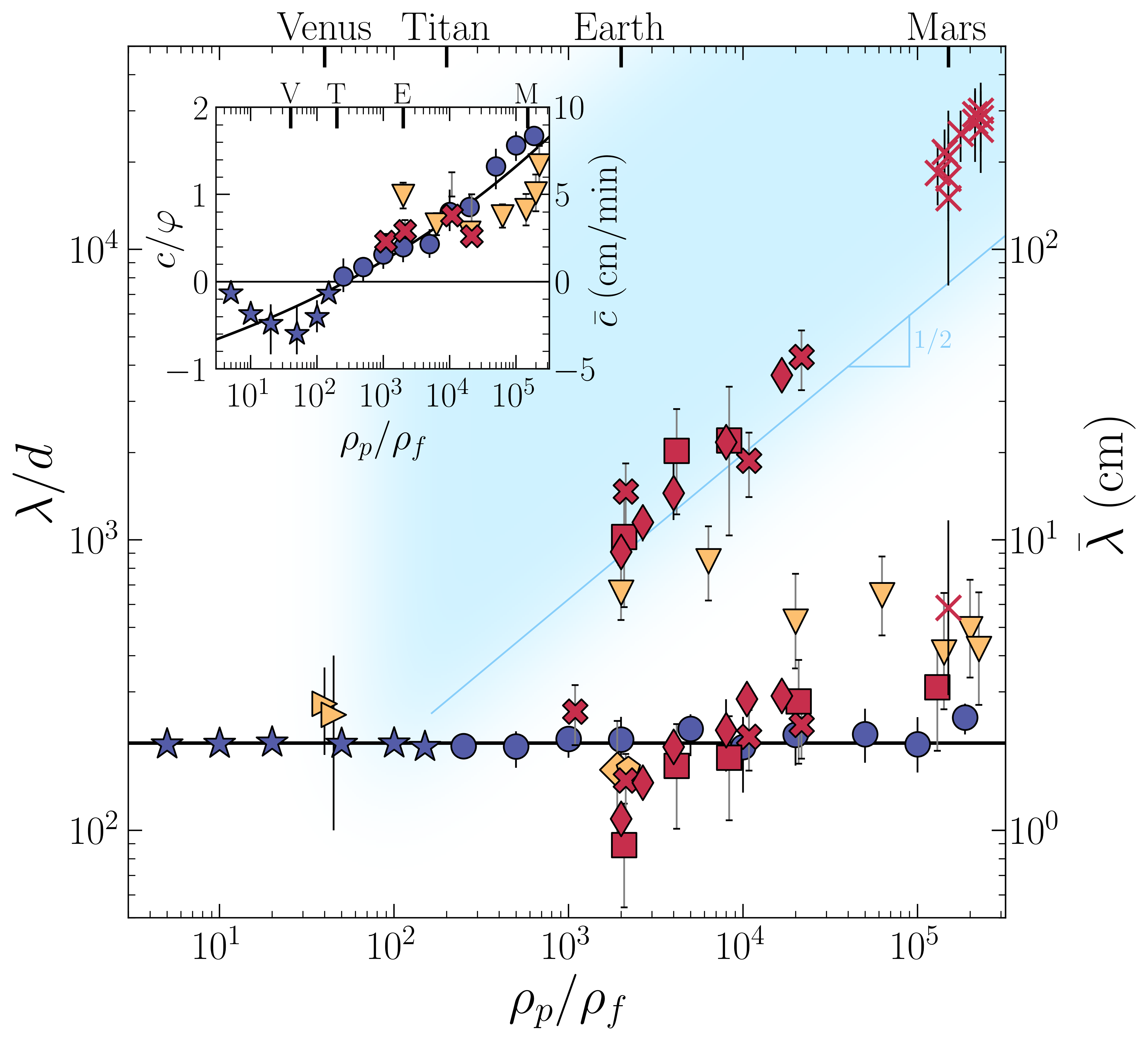}
    \caption{Observed and simulated ripple wavelengths $\lambda$ rescaled by grain size $d$ for different grain-fluid density ratios $\rho_p/\rho_f$---representing a range of planetary environments (different planets are marked on the top axis). The right axis $\bar\lambda$ represents $\lambda/d$ scaled by $d=100\ \mu$m. Simulations are shown in purple symbols. Observations that report two coexisting bedforms are shown in red: \citeA{miller1987wind} ($\lozenge$), \citeA{yizhaq2024coevolving} ($\square$), \citeA{alvarez2025ripples} (\rotatebox[origin=c]{45}{\contour{black}{\color{white}{\ding{58}}}}), \citeA{lapotre2016large} ($\times$). Observations that report one single bedform are shown in gold: \citeA{greeley1984microdunes} (\rotatebox[origin=c]{-90}{$\triangle$}; smallest bedforms reported), \citeA{andreotti2006aeolian} (\rotatebox[origin=c]{45}{$\square$}), \citeA{andreotti2021lower} (\rotatebox[origin=c]{180}{$\triangle$}). All observational data are for small grain sizes $d\sim100\ \mu$m at wind speeds near threshold. Grain-scale ripple simulations ($\Circle$ for downstream and $\star$ for upstream migration) for all simulated grain sizes close to threshold ($Q_*<0.1$) show $\lambda \simeq 200d$ (Fig. \ref{fig3}b). Vertical bars denote observed wavelength ranges and thin capped error bars are reported standard deviations. The shaded region represents the predicted hydrodynamic ripple regime bounded by the initial wavelength approximation $\lambda_\text{hydro}=1000\nu/u_*$ (blue line). The hydrodynamic ripple regime vanishes for hydrodynamically rough flows when $\rho_p/\rho_f\lesssim10^2$ for $d\simeq 100\ \mu$m (see \ref{app: HR scaling}). Inset shows rescaled ripple speeds $c/\varphi$ from simulations (averages and ranges for $d\lesssim 200\ \mu$m, $\Circle$ and $\star$) compared with impact ripple speeds measured by \citeA{alvarez2025ripples} (\rotatebox[origin=c]{45}{\contour{black}{\color{white}{\ding{58}}}})---using the DEM approximation $\varphi/\sqrt{gd}\simeq(\Theta-\Theta_\text{th})/2$. Data from \citeA{andreotti2021lower} (\rotatebox[origin=c]{180}{$\triangle$}) also shown. The right axis in the inset $\bar{c}$ scale is scaled by $\varphi=5$ cm/min.}
    \label{fig4}
\end{figure}

\section{Ripple Scaling Across Planetary Environments}\label{sec: data comp}

The main control parameters associated with impact ripple dynamics are the fluid density $\rho_f$, grain size $d$, and wind speed $u_*$. Many wind-tunnel and field observations have studied ripple sizes and speeds in this three-dimensional parameter space. Several experiments \cite{yizhaq2024coevolving,alvarez2025ripples,miller1987wind} have reported two coevolving ripple scales, hypothesized as impact and hydrodynamic ripples. Here we compare simulated impact ripples to a wide range of experimental and field observations of aeolian bedforms across environments. To better distinguish potential hydrodynamic ripples from impact ripples in the data, we introduce a hydrodynamic-ripple scaling derived from the framework of \citeA{duran2019unified}, which predicts $\lambda_\text{hydro}\propto d_\mu \sqrt{\rho_p/\rho_f} \sqrt{d_\mu/d}$ near the transport threshold (see \ref{app: HR scaling}).

\subsection{Fluid density}

We approximate the conditions on other planetary bodies in the solar system primarily by changing the grain-fluid density ratio $\rho_p/\rho_f$. We find that ripples form under all possible conditions, from Martian-like ($\rho_p/\rho_f\sim 10^5$), to Earth ($\rho_p/\rho_f\sim 2000$), Titan ($\rho_p/\rho_f\sim 200$) and Venus ($\rho_p/\rho_f\sim 50$). Surprisingly, we also find impact ripples under water-like bedload transport ($\rho_p/\rho_f\lesssim 10$). 

For low-intensity transport (`saltation' regime, $Q_*\lesssim0.1$), impact ripple wavelengths remain close to $\lambda\simeq 200d$ as $\rho_p/\rho_f$ varies by five orders of magnitude (Fig.~\ref{fig4}), as predicted by the theory \cite{lester2025emergence}, and is consistent with the smallest bedforms reported in experiments that observe two coexisting ripple scales \cite{yizhaq2024coevolving,alvarez2025ripples,miller1987wind}. By contrast, the wavelength of the larger ripples increases with $\rho_p/\rho_f$ as predicted by the hydrodynamic ripple scaling $\lambda_\text{hydro}\propto d_\mu \sqrt{\rho_p/\rho_f}$, supporting their interpretation as a distinct hydrodynamic branch. Among studies reporting only a single bedform scale (gold markers, Fig.~\ref{fig4}), several datasets remain compatible with the impact-ripple range, whereas the low-pressure wind-tunnel experiments of \citeA{andreotti2021lower} report ripple wavelengths that remain nearly fixed across atmospheric pressures but are 2-4 times larger than DEM impact ripples and other wind tunnel experiments in similar conditions.

While impact-ripple wavelengths do not vary much with density ratio $\rho_p/\rho_f$, ripple migration speeds $c$ vary strongly across environments---from positive for Mars and Earth-like conditions, near-stationary for Titan-like conditions, and negative (antiripples) for Venus-like conditions (Fig.~\ref{fig4} inset). This strong environmental sensitivity is consistent with Equation~\ref{eq: c}, because the average hoplength $\langle\ell\rangle$ decreases continuously with the density ratio $\rho_p/\rho_f$, eventually crossing the critical hoplength for the ripple-antiripple transition around $\rho_p/\rho_f\sim 10^2$ (Fig.~\ref{fig4} inset). Although no observations yet reach sufficiently low density ratios to test the antiripple prediction directly, the available impact-ripple speed measurements from \citeA{alvarez2025ripples}, and possibly \citeA{andreotti2021lower}, are broadly consistent with DEM simulations and analytical results (Fig.~\ref{fig4} inset).

\begin{figure}[t!]
    \noindent\includegraphics[width=.7\textwidth]{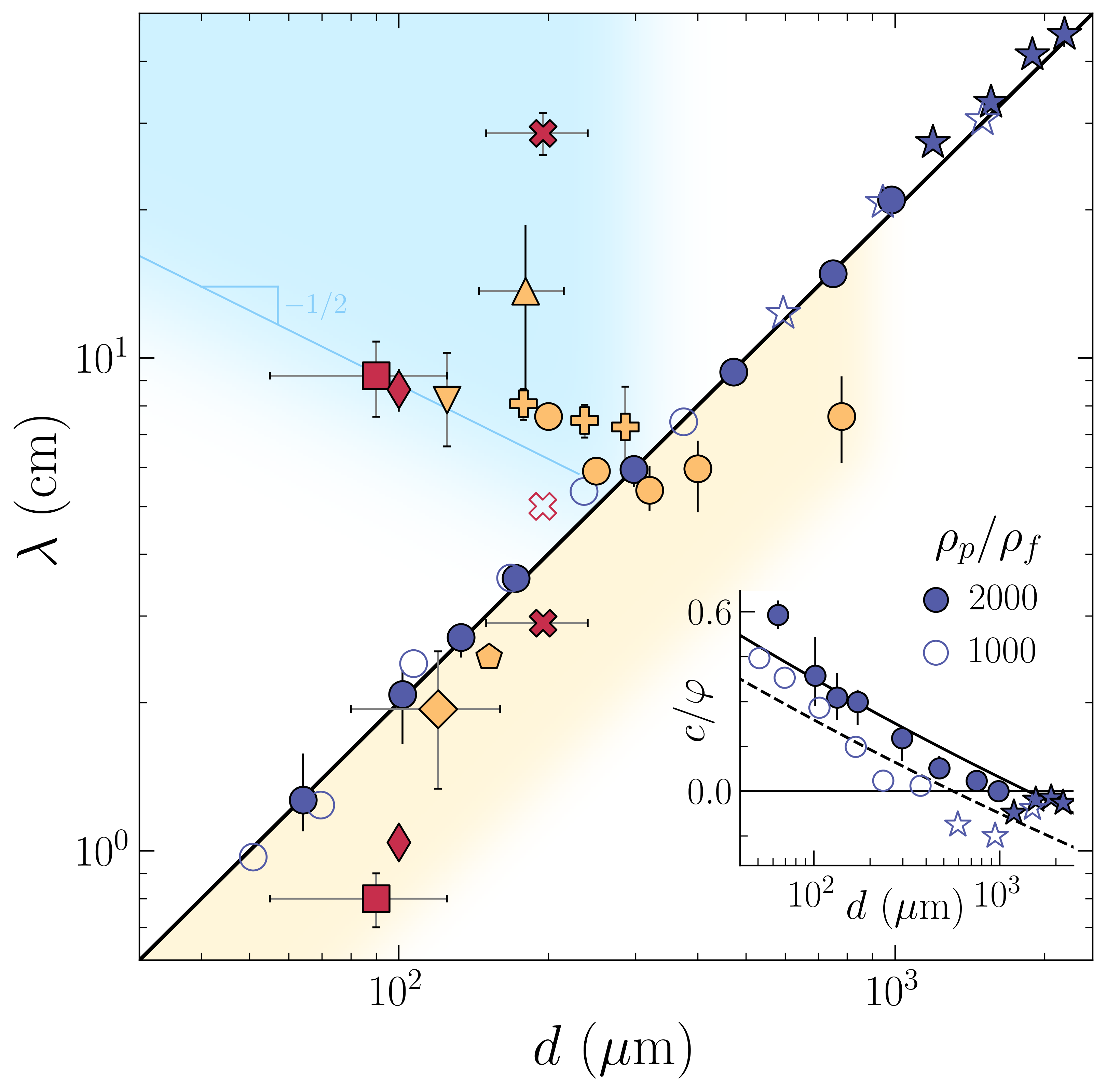}
    \caption{Observed and simulated ripple wavelengths $\lambda$ versus grain size $d$ for Earth-like conditions $\rho_p/\rho_f\simeq 2000$. Wind tunnel observations that report two coexisting bedform wavelengths (red): \citeA{miller1987wind} ($\lozenge$), \citeA{yizhaq2024coevolving} ($\square$), \citeA{alvarez2025ripples} (\rotatebox[origin=c]{45}{\contour{black}{\color{white}{\ding{58}}}}). Observations that report a single bedform wavelength (gold): wind tunnel (\rotatebox[origin=c]{45}{$\square$}; average of slowest two wind speeds) and field ($\triangle$; average and range for $u_*/\sqrt{gd}<10$) observations from \citeA{andreotti2006aeolian}, \citeA{andreotti2021lower} (\rotatebox[origin=c]{180}{$\triangle$}), \citeA{cheng2018experimental} (\pentagon; slowest wind speed), \citeA{schmerler2016experimental} (\contour{black}{\color{white}{\ding{58}}}; slowest wind speeds for each $d$), \citeA{walker1981experimental} ($\bigcirc$; averages and ranges for $u_*/\sqrt{gd}<10$ for each $d$). DEM simulations are shown in purple for $Q_*<0.1$ with antiripples marked as stars. Open symbols are for $\rho_p/\rho_f=1000$ representing light grains---similar density ratio of crushed nut shells used in \citeA{alvarez2025ripples} (\rotatebox[origin=c]{45}{\contour{black}{\color{white}{\ding{58}}}}). Markers with bars show the range of ripples observed and capped error bars show standard deviations. Hydrodynamic ripple region (blue shade) is bounded by the initial wavelength approximation $\lambda_\text{hydro}=1000\nu/u_*$ (blue line) and the hydrodynamic roughness transition at $d\gtrsim 200\ \mu$m for $\rho_p/\rho_f=2000$ (see \ref{app: HR scaling}). Observed ripple wavelengths outside of the hydrodynamic regime fall within a factor of $2$ (gold shade) of DEM simulated ripples $\lambda=200d$ (black line). Inset shows DEM measured ripple speeds $c/\varphi$ and predictions (Eq.~\ref{eq: c}). Note the antiripple transition ($c<0$; stars) occurs for $d\gtrsim 500\ \mu$m for $\rho_p/\rho_f=1000$ and $d\gtrsim 1000\ \mu$m for $\rho_p/\rho_f=2000$.}
    \label{fig5}
\end{figure}

\subsection{Grain size}

Figure~\ref{fig5} shows the relation between ripple wavelengths and grain size on Earth, comparing simulated impact ripples with wind tunnel and field observations for wind speeds relatively close to threshold (`saltation' regime). For Earth-like conditions ($\rho_p/\rho_f=2000$), experiments show two coexisting ripple sizes for relatively fine sand ($d\lesssim200\ \mu$m), whereas there seems to be a single ripple size for coarser sand. The larger ripples have been described as hydrodynamic in origin, with a wavelength roughly scaling with the viscous length $\nu/u_*$ \cite{duran2019unified} and therefore inversely with the square root of the grain size: $\lambda_\text{hydro}\propto d_\mu\sqrt{d_\mu/d}$ (\ref{app: HR scaling}). The smaller ripples are consistent with impact ripples, with a wavelength scaling with the grain size, and are well described by the DEM simulations, which provide an upper bound for their wavelength~\cite{yizhaq2024coevolving}. For grain sizes $d\gtrsim 200\ \mu$m, observational data cluster near $\lambda \sim 5$–10 cm. We interpret this region as an overlap regime in which impact and hydrodynamic instabilities compete to select a similar wavelength, thus making ripple formation mechanism more ambiguous. 

The grain-size dependence also yields a direct experimental test of the antiripple prediction. Because the average hoplength increases sublinearly with grain size, the DEM predicts a ripple–antiripple transition for sufficiently large grains: around $d\gtrsim 1000$ $\mu$m for Earth-like density ratios and around $d\gtrsim 500$ $\mu$m for lighter grains ($\rho_p/\rho_f=1000$; Fig.~\ref{fig5} inset). 

\begin{figure}[t!]
    \noindent\includegraphics[width=\textwidth]{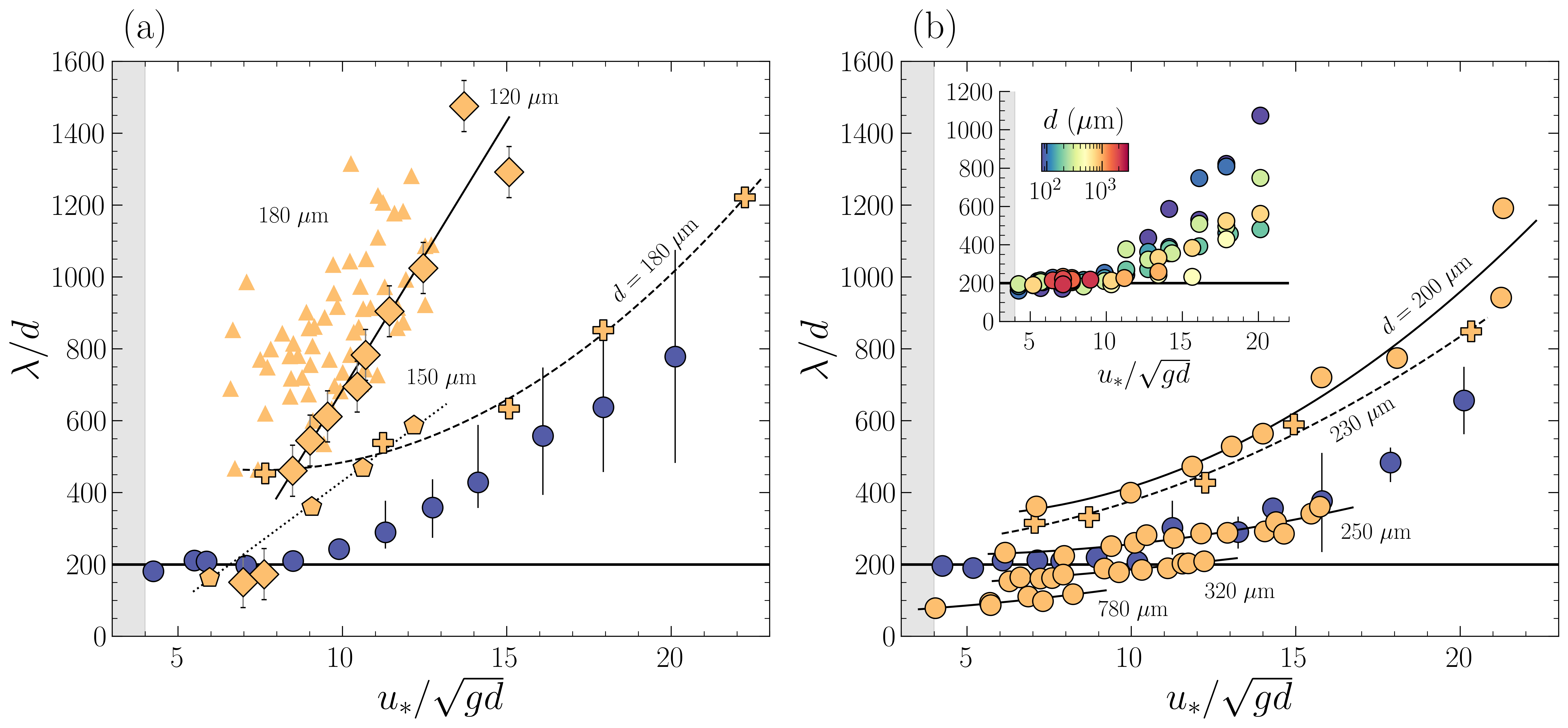}
    \caption{Ripple size $\lambda$ as a function of shear velocity $u_*$ on Earth. Panel (a) shows wavelength data for $d<200\ \mu$m and (b) shows data for $d\geq200\ \mu$m. Wind tunnel and field data are gold markers and DEM simulations are purple markers (for $\rho_p/\rho_f=2000$). Observational data symbols correspond to those in Figures~\ref{fig4} and \ref{fig5}. Reported grain sizes are marked with the observational data as well as trend lines to aid the eye. In (a) DEM simulated wavelengths $\lambda$ are averaged for all simulations with $d< 200\ \mu$m with observed ranges shown by the black bars. And (b) shows averaged simulations for $d>200\ \mu$m. Inset in (b) shows DEM simulations for all grain sizes (marker colors). In all panels the gray boundary marks an approximate threshold shear velocity $u_\text{th}=\sqrt{\Theta_\text{th}(\rho_p/\rho_f-1)gd}\simeq 4\sqrt{gd}$ for $\Theta_\text{th}=0.008$ \cite{pahtz2018cessation}.}
    \label{fig6}
\end{figure}

\begin{figure}[t!]
    \noindent\includegraphics[width=\textwidth]{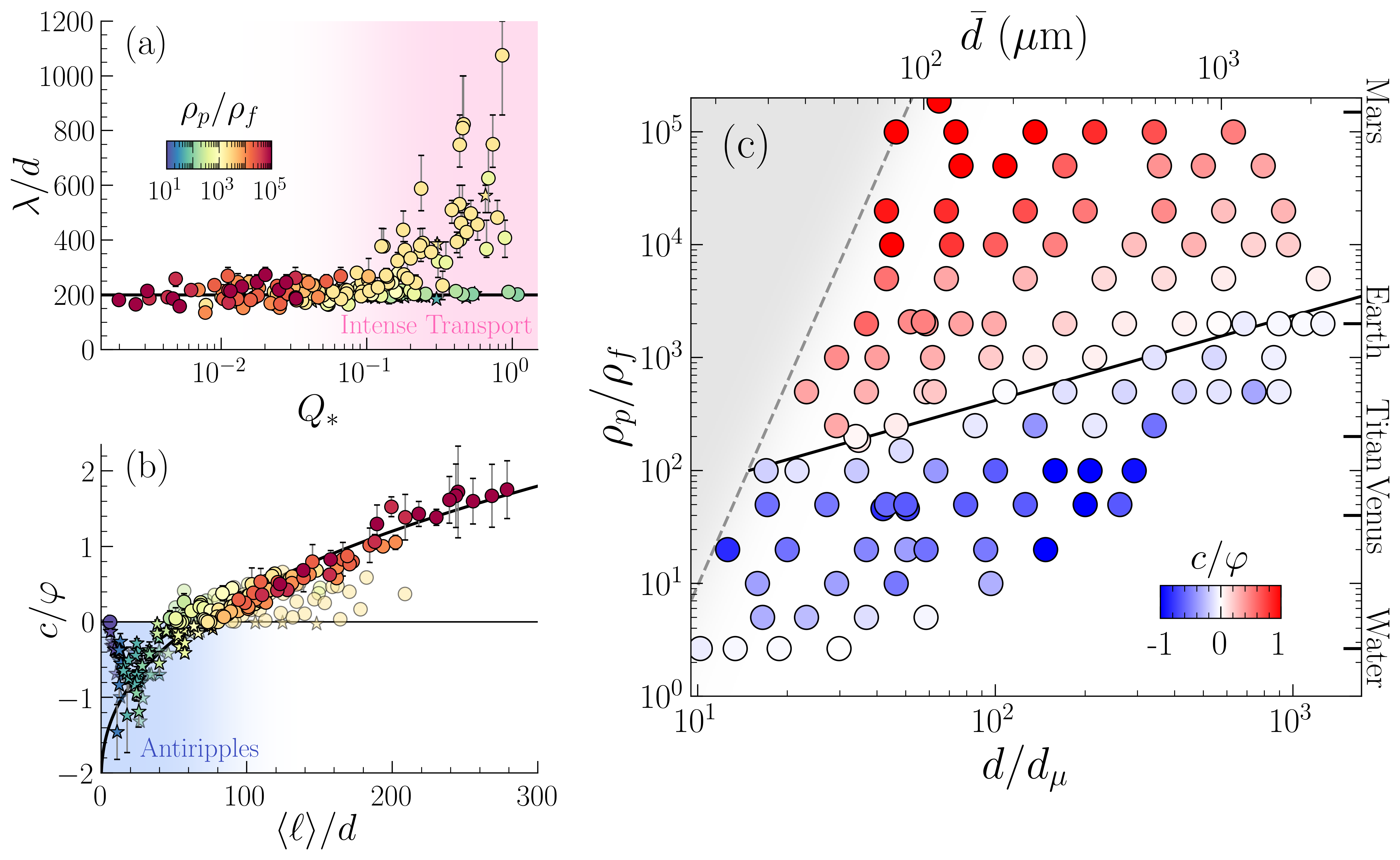}
    \caption{Summary of the scaling of simulated impact ripples and ripple-antiripple transition. (a) Rescaled wavelengths $\lambda/d$ versus rescaled transport rate $Q_*$ (Eq.~\ref{eq: Q*}). Symbol color denotes the density ratio. Shaded region highlights the `collisional' high-intensity transport regime. (b) Propagation speeds $c$ rescaled by the vertical flux at the surface $\varphi$ as a function of rescaled average hoplength $\langle\ell\rangle/d$ (symbol colors same as (a)). Color intensity differentiates `saltation' (intense color) from `collisional' (faded color) regimes. Solid lines in both (a) and (b) are predictions from \citeA{lester2025emergence} for the `saltation' regime ($\lambda=200d$ and Eq.~\ref{eq: c}). (c) Ripple/antiripple phase space in terms of the density ratio $\rho_p/\rho_f$ and rescaled grain size $d/d_\mu$. Top axis $\bar{d}$ is obtained using $d_\mu=1.74\ \mu$m representing Earth-like conditions. Right axis marks the approximate density ratios of specific planets. Symbols are DEM simulations close to the transport threshold, where the scaling $\lambda\simeq 200 d$ holds. Markers are colored by the rescaled ripple speed $c/\varphi$. The grey region/dashed line marks the regime where sediment transport could transition to suspension at the transport threshold $u_\text{th}$ and the model is not applicable~\cite{pahtz2023scaling}. The solid black line denotes the approximate ripple-antiripple transition ($c=0$) when $\langle\ell\rangle=\langle\ell\rangle_c\simeq67d$ (Eq.~\ref{eq: c}), which when using the DEM scaling for $\langle\ell\rangle$ reduces to the relation $\rho_p/\rho_f \simeq 13 (d/d_\mu)^{0.75}$.}
    \label{fig7}
\end{figure}

\subsection{Wind speed}

A major focus of wind-tunnel and field studies on Earth has been the dependence of ripple wavelengths on the wind speed, typically characterized by the shear velocity $u_*$~\cite{andreotti2006aeolian,cheng2018experimental,schmerler2016experimental,walker1981experimental}. Interpreting the reported $\lambda(u_*)$ trends, however, is complicated by strong variability between datasets and by clear differences among grain sizes $d$ (Fig.~\ref{fig6}). In particular, experiments with fine grains (Fig.~\ref{fig6}a) generally show a stronger increase of $\lambda$ with $u_*$ than experiments with coarser grains (Fig.~\ref{fig6}b). This separation by grain size is important because fine-grain experiments are also the conditions under which impact and hydrodynamic bedforms may coexist (Fig.~\ref{fig5}), making bedform identification less straightforward.

Although the DEM simulations reproduced the same broad $\lambda(u_*)$ trend of wind-tunnel experiments, the observed wavelengths for fine grains were consistently underestimated by the simulations. This could result from the hydrodynamic origin of most reported ripple data for grain sizes below $200\ \mu$m (Fig.~\ref{fig5}). 

On Earth, simulated impact-ripple wavelengths remain close to $200d$ for wind shear velocities below $u_*\lesssim 10\sqrt{gd}$, increasing with $u_*$ for $u_*\gtrsim 10\sqrt{gd}$ (Fig.~\ref{fig6}). As expected, the condition $u_*\simeq 10\sqrt{gd}$ is consistent with the value $Q_* \simeq 0.1$ separating `saltation' from `collisional' regimes. The increase with the shear velocity follows from the scaling of impact ripple wavelengths with the average hoplength $\langle\ell\rangle$ in the `collisional' regime, and the relation between $\langle\ell\rangle$ and $u_*$ \cite{duran2014direct}. Interestingly, simulations show that this increase is more pronounced for fine grains (Fig.~\ref{fig6} inset), qualitatively consistent with the experimental trends.

\section{Discussion and Conclusions}

Aeolian ripples across planetary conditions have proven to be highly variable in wind tunnel experiments \cite{andreotti2006aeolian,andreotti2021lower,greeley1984microdunes,miller1987wind,alvarez2025ripples, yizhaq2024coevolving,cheng2018experimental,schmerler2016experimental,walker1981experimental} and in the field \cite{lapotre2016large,andreotti2006aeolian}. The lack of clarity in observed scaling relationships of aeolian bedforms on Earth may be explained by the discovery that, for fine grains, two scale-separated aeolian bedforms can coexist simultaneously \cite{alvarez2025ripples,yizhaq2024coevolving,miller1987wind}---small impact ripples and larger proposed hydrodynamic ripples. Here we have shown that direct numerical simulations of impact ripples quantitatively reproduce the smallest ripples reported in dual bedform observations \cite{alvarez2025ripples,yizhaq2024coevolving,miller1987wind,lapotre2016large}, while the larger reported ripples are consistent with hydrodynamic-ripple scaling---although more experimental work is needed in this regard, especially with varying grain sizes and wind speeds.

Simulations show that impact-ripple dynamics can be roughly organized by transport intensity regimes. For low-intensity transport, $Q_*\lesssim 0.1$, ripple wavelengths remain relatively constant around $\lambda\simeq 200d$ (Fig.~\ref{fig7}a). For intense transport, $Q_*\gtrsim 0.1$, wavelengths generally tend to increase with transport scales and wind speed. This scaling separation qualitatively resolves tension between wavelength-selection models that predict grain-size-selected $\lambda\propto d$ \cite{lester2025emergence} and hoplength-selected $\lambda\propto \langle\ell\rangle$ \cite{duran2014direct} ripple sizes by suggesting that they both apply in different transport regimes.

The most notable variability in simulated impact ripple dynamics is the ripple propagation speeds which vary considerably with atmospheric density, grain size and wind speed (Fig.~\ref{fig7}b,c). The scaling of the ripple speeds arises from the transport rate and the average hoplength (Eq.~\ref{eq: c}; Fig.~\ref{fig7}b), with the notable prediction that below a critical hoplength scale (set by a critical fluid density and grain size; Fig.~\ref{fig7}c) impact ripples propagate against the wind as antiripples \cite{lester2025emergence}. The phase space in Figure~\ref{fig7}c for impact ripples at low transport intensity ($\lambda\simeq 200d$) summarizes the ripple-antiripple regimes scaling across planetary conditions, and motivates future experiments on where to probe to validate the existence of antiripples.

In this paper we have outlined how numerical simulations and analytical predictions \cite{lester2025emergence} shed light on impact ripple scaling across planetary environments. This work provides a range of predictions to be tested in future experiments, including the simplified scaling of impact ripple sizes with grain size and the existence of antiripples. Further experimental work is also needed to test for the coexistence of impact-driven and hydrodynamically driven aeolian bedforms.

\appendix  

\section{Hydrodynamic Ripples}\label{app: HR scaling}

The largest ripples reported in some wind-tunnel and field observations \cite{yizhaq2024coevolving,alvarez2025ripples,lapotre2016large} are proposed to be hydrodynamic in nature. This suggests that the predictive hydrodynamic bedform model of \citeA{duran2019unified} can be used to test this prediction. From this model, hydrodynamic bedforms sizes fall in the wavelength corridor  $500\nu/u_*\lesssim \lambda_\text{hydro}  \lesssim10^4\nu/u_*$. The predicted initial wavelength is determined by the saturation length $\ell_\text{sat}$ (the transient lag between flux and shear stress) and scales as $\lambda_\text{hydro}u_*/\nu\propto (\ell_\text{sat}u_*/\nu)^\beta$ for $\beta\simeq 0.6$. However, on Earth $\ell_\text{sat}\sim 1/2$ m, which is too large to set the observed hydrodynamic ripple wavelengths. \citeA{yizhaq2024coevolving} proposed a phenomenological sub-scale saturation length $\ell_\text{sat}^\text{sub}\sim 1\text{ cm}\ll \ell_\text{sat}$ associated with a layered transient relaxation response of the flux to changes in shear stress. The existence of this sub-scale saturation length suggests a fundamental gap in our understanding of how transport responds to changes in bed shear stress. Because of this uncertainty, we assume initial hydrodynamic ripple wavelengths scale with the viscous length $\nu/u_*$ and are of order 
\begin{equation}
    \label{eq: lam hydro}
    \lambda_\text{hydro}\sim 1000 \frac{\nu}{u_*},
\end{equation}
as reported in observations \cite{yizhaq2024coevolving,alvarez2025ripples,lapotre2016large}. 

For hydrodynamic ripple scaling predictions across environments, we assume the wind speed is close to the cessation/dynamic threshold $u_*\sim u_\text{th}$ and take the threshold shear velocity $u_\text{th}$ as set by the constant Shields number $\Theta_\text{th}=0.008$ (as reported in observations \cite{pahtz2018cessation}). This leads to the scaling 
\begin{equation}
    \label{eq: lam HR scaling}
    \frac{\lambda_\text{hydro}}{d_\mu}\propto \sqrt{\frac{\rho_p}{\rho_f}\frac{d_\mu}{d}}
\end{equation}
with proportionality constant $ 1000/\sqrt{\Theta}$. Because of the uncertainty in the sub-scale saturation length dynamics, we do not compare hydrodynamic predictions for varying wind speeds $u_*$. Finally, \citeA{duran2019unified} show that hydrodynamic bedform wavelengths transition to dune scale when $u_*d_\text{eff}/\nu\gtrsim20$ (the `rough' regime) where the effective grain size $d_\text{eff}$ is set by the transport-layer roughness height, which increases with wind speed above threshold $d_\text{eff}/d\sim \exp(-7(u_\text{th}/u_*-1))$. We choose the arbitrary value $u_*/u_\text{th}=1.3$, which corresponds to $d_\text{eff}\simeq 5d$ and $\sqrt{\Theta}\simeq 0.12$. (Variations in $u_*/u_\text{th}$ do not change the qualitative results.)





\section*{Open Research Statement}
The DEM simulation data shown in Figures~\ref{fig3}--\ref{fig7}, including the input parameters of each simulation, are available at Zenodo \cite{lester2026data}. The DEM model was developed by \citeA{duran2012numerical}, where it is described in detail, and has been used in several previous studies \cite<e.g.,>{duran2014direct,pahtz2020unification,pahtz2023scaling,lester2025emergence}. The model code is not publicly archived but is available from O. Duran upon request. All non-DEM data shown in Figures~\ref{fig4}--\ref{fig6} are from the publications cited in the figure captions.

\section*{Conflict of Interest Statement}
The authors declare there are no conflicts of interest for this manuscript.


%
\bibliography{references}

\begin{thebibliography}{}

\bibitem [\protect \citeauthoryear {%
Alvarez%
, Lap{\^o}tre%
, Swann%
\BCBL {}\ \BBA {} Ewing%
}{%
Alvarez%
\ \protect \BOthers {.}}{%
{\protect \APACyear {2025}}%
}]{%
alvarez2025ripples}
\APACinsertmetastar {%
alvarez2025ripples}%
\begin{APACrefauthors}%
Alvarez, C\BPBI A.%
, Lap{\^o}tre, M\BPBI G\BPBI A.%
, Swann, C.%
\BCBL {}\ \BBA {} Ewing, R\BPBI C.%
\end{APACrefauthors}%
\unskip\
\newblock
\APACrefYearMonthDay{2025}{}{}.
\newblock
{\BBOQ}\APACrefatitle {Ripples formed in low-pressure wind tunnels suggest
  {M}ars’s large windblown ripples are not impact ripples} {Ripples formed in
  low-pressure wind tunnels suggest {M}ars’s large windblown ripples are not
  impact ripples}.{\BBCQ}
\newblock
\APACjournalVolNumPages{Nature Communications}{16}{1}{2945}.
\newblock
\begin{APACrefDOI} \doi{10.1038/s41467-025-58140-7} \end{APACrefDOI}
\PrintBackRefs{\CurrentBib}

\bibitem [\protect \citeauthoryear {%
Anderson%
}{%
Anderson%
}{%
{\protect \APACyear {1987}}%
}]{%
anderson1987theoretical}
\APACinsertmetastar {%
anderson1987theoretical}%
\begin{APACrefauthors}%
Anderson, R\BPBI S.%
\end{APACrefauthors}%
\unskip\
\newblock
\APACrefYearMonthDay{1987}{}{}.
\newblock
{\BBOQ}\APACrefatitle {A theoretical model for aeolian impact ripples} {A
  theoretical model for aeolian impact ripples}.{\BBCQ}
\newblock
\APACjournalVolNumPages{Sedimentology}{34}{5}{943--956}.
\newblock
\begin{APACrefDOI} \doi{10.1111/j.1365-3091.1987.tb00814.x} \end{APACrefDOI}
\PrintBackRefs{\CurrentBib}

\bibitem [\protect \citeauthoryear {%
Andreotti%
, Claudin%
, Iversen%
, Merrison%
\BCBL {}\ \BBA {} Rasmussen%
}{%
Andreotti%
\ \protect \BOthers {.}}{%
{\protect \APACyear {2021}}%
}]{%
andreotti2021lower}
\APACinsertmetastar {%
andreotti2021lower}%
\begin{APACrefauthors}%
Andreotti, B.%
, Claudin, P.%
, Iversen, J\BPBI J.%
, Merrison, J\BPBI P.%
\BCBL {}\ \BBA {} Rasmussen, K\BPBI R.%
\end{APACrefauthors}%
\unskip\
\newblock
\APACrefYearMonthDay{2021}{}{}.
\newblock
{\BBOQ}\APACrefatitle {A lower-than-expected saltation threshold at {M}artian
  pressure and below} {A lower-than-expected saltation threshold at {M}artian
  pressure and below}.{\BBCQ}
\newblock
\APACjournalVolNumPages{Proceedings of the National Academy of
  Sciences}{118}{5}{e2012386118}.
\newblock
\begin{APACrefDOI} \doi{10.1073/pnas.2012386118} \end{APACrefDOI}
\PrintBackRefs{\CurrentBib}

\bibitem [\protect \citeauthoryear {%
Andreotti%
, Claudin%
\BCBL {}\ \BBA {} Pouliquen%
}{%
Andreotti%
\ \protect \BOthers {.}}{%
{\protect \APACyear {2006}}%
}]{%
andreotti2006aeolian}
\APACinsertmetastar {%
andreotti2006aeolian}%
\begin{APACrefauthors}%
Andreotti, B.%
, Claudin, P.%
\BCBL {}\ \BBA {} Pouliquen, O.%
\end{APACrefauthors}%
\unskip\
\newblock
\APACrefYearMonthDay{2006}{}{}.
\newblock
{\BBOQ}\APACrefatitle {Aeolian sand ripples: Experimental study of fully
  developed states} {Aeolian sand ripples: Experimental study of fully
  developed states}.{\BBCQ}
\newblock
\APACjournalVolNumPages{Physical Review Letters}{96}{2}{028001}.
\newblock
\begin{APACrefDOI} \doi{10.1103/physrevlett.96.028001} \end{APACrefDOI}
\PrintBackRefs{\CurrentBib}

\bibitem [\protect \citeauthoryear {%
Bagnold%
}{%
Bagnold%
}{%
{\protect \APACyear {1941}}%
}]{%
Bagnold1941}
\APACinsertmetastar {%
Bagnold1941}%
\begin{APACrefauthors}%
Bagnold, R\BPBI A.%
\end{APACrefauthors}%
\unskip\
\newblock
\APACrefYear{1941}.
\newblock
\APACrefbtitle {The Physics of Blown Sand and Desert Dunes} {The physics of
  blown sand and desert dunes}.
\newblock
\APACaddressPublisher{London}{Methuen}.
\PrintBackRefs{\CurrentBib}

\bibitem [\protect \citeauthoryear {%
Charru%
, Andreotti%
\BCBL {}\ \BBA {} Claudin%
}{%
Charru%
\ \protect \BOthers {.}}{%
{\protect \APACyear {2013}}%
}]{%
charru2013sand}
\APACinsertmetastar {%
charru2013sand}%
\begin{APACrefauthors}%
Charru, F.%
, Andreotti, B.%
\BCBL {}\ \BBA {} Claudin, P.%
\end{APACrefauthors}%
\unskip\
\newblock
\APACrefYearMonthDay{2013}{}{}.
\newblock
{\BBOQ}\APACrefatitle {Sand ripples and dunes} {Sand ripples and dunes}.{\BBCQ}
\newblock
\APACjournalVolNumPages{Annual Review of Fluid Mechanics}{45}{1}{469--493}.
\newblock
\begin{APACrefDOI} \doi{10.1146/annurev-fluid-011212-140806} \end{APACrefDOI}
\PrintBackRefs{\CurrentBib}

\bibitem [\protect \citeauthoryear {%
Cheng%
\ \protect \BOthers {.}}{%
Cheng%
\ \protect \BOthers {.}}{%
{\protect \APACyear {2018}}%
}]{%
cheng2018experimental}
\APACinsertmetastar {%
cheng2018experimental}%
\begin{APACrefauthors}%
Cheng, H.%
, Liu, C.%
, Li, J.%
, Liu, B.%
, Zheng, Z.%
, Zou, X.%
\BDBL {}Fang, Y.%
\end{APACrefauthors}%
\unskip\
\newblock
\APACrefYearMonthDay{2018}{}{}.
\newblock
{\BBOQ}\APACrefatitle {Experimental study of aeolian sand ripples in a wind
  tunnel} {Experimental study of aeolian sand ripples in a wind tunnel}.{\BBCQ}
\newblock
\APACjournalVolNumPages{Earth Surface Processes and
  Landforms}{43}{1}{312--321}.
\newblock
\begin{APACrefDOI} \doi{10.1002/esp.4246} \end{APACrefDOI}
\PrintBackRefs{\CurrentBib}

\bibitem [\protect \citeauthoryear {%
Csah{\'o}k%
, Misbah%
, Rioual%
\BCBL {}\ \BBA {} Valance%
}{%
Csah{\'o}k%
\ \protect \BOthers {.}}{%
{\protect \APACyear {2000}}%
}]{%
csahok2000dynamics}
\APACinsertmetastar {%
csahok2000dynamics}%
\begin{APACrefauthors}%
Csah{\'o}k, Z.%
, Misbah, C.%
, Rioual, F.%
\BCBL {}\ \BBA {} Valance, A.%
\end{APACrefauthors}%
\unskip\
\newblock
\APACrefYearMonthDay{2000}{}{}.
\newblock
{\BBOQ}\APACrefatitle {Dynamics of aeolian sand ripples} {Dynamics of aeolian
  sand ripples}.{\BBCQ}
\newblock
\APACjournalVolNumPages{The European Physical Journal E}{3}{}{71--86}.
\newblock
\begin{APACrefDOI} \doi{10.1007/s101890070043} \end{APACrefDOI}
\PrintBackRefs{\CurrentBib}

\bibitem [\protect \citeauthoryear {%
Dur{\'a}n%
, Andreotti%
\BCBL {}\ \BBA {} Claudin%
}{%
Dur{\'a}n%
\ \protect \BOthers {.}}{%
{\protect \APACyear {2012}}%
}]{%
duran2012numerical}
\APACinsertmetastar {%
duran2012numerical}%
\begin{APACrefauthors}%
Dur{\'a}n, O.%
, Andreotti, B.%
\BCBL {}\ \BBA {} Claudin, P.%
\end{APACrefauthors}%
\unskip\
\newblock
\APACrefYearMonthDay{2012}{}{}.
\newblock
{\BBOQ}\APACrefatitle {Numerical simulation of turbulent sediment transport,
  from bed load to saltation} {Numerical simulation of turbulent sediment
  transport, from bed load to saltation}.{\BBCQ}
\newblock
\APACjournalVolNumPages{Physics of Fluids}{24}{10}{103306}.
\newblock
\begin{APACrefDOI} \doi{10.1063/1.4757662} \end{APACrefDOI}
\PrintBackRefs{\CurrentBib}

\bibitem [\protect \citeauthoryear {%
Dur{\'a}n%
, Claudin%
\BCBL {}\ \BBA {} Andreotti%
}{%
Dur{\'a}n%
\ \protect \BOthers {.}}{%
{\protect \APACyear {2011}}%
}]{%
duran2011aeolian}
\APACinsertmetastar {%
duran2011aeolian}%
\begin{APACrefauthors}%
Dur{\'a}n, O.%
, Claudin, P.%
\BCBL {}\ \BBA {} Andreotti, B.%
\end{APACrefauthors}%
\unskip\
\newblock
\APACrefYearMonthDay{2011}{}{}.
\newblock
{\BBOQ}\APACrefatitle {On aeolian transport: Grain-scale interactions,
  dynamical mechanisms and scaling laws} {On aeolian transport: Grain-scale
  interactions, dynamical mechanisms and scaling laws}.{\BBCQ}
\newblock
\APACjournalVolNumPages{Aeolian Research}{3}{3}{243--270}.
\newblock
\begin{APACrefDOI} \doi{10.1016/j.aeolia.2011.07.006} \end{APACrefDOI}
\PrintBackRefs{\CurrentBib}

\bibitem [\protect \citeauthoryear {%
Dur{\'a}n%
, Claudin%
\BCBL {}\ \BBA {} Andreotti%
}{%
Dur{\'a}n%
\ \protect \BOthers {.}}{%
{\protect \APACyear {2014}}%
}]{%
duran2014direct}
\APACinsertmetastar {%
duran2014direct}%
\begin{APACrefauthors}%
Dur{\'a}n, O.%
, Claudin, P.%
\BCBL {}\ \BBA {} Andreotti, B.%
\end{APACrefauthors}%
\unskip\
\newblock
\APACrefYearMonthDay{2014}{}{}.
\newblock
{\BBOQ}\APACrefatitle {Direct numerical simulations of aeolian sand ripples}
  {Direct numerical simulations of aeolian sand ripples}.{\BBCQ}
\newblock
\APACjournalVolNumPages{Proceedings of the National Academy of
  Sciences}{111}{44}{15665--15668}.
\newblock
\begin{APACrefDOI} \doi{10.1073/pnas.1413058111} \end{APACrefDOI}
\PrintBackRefs{\CurrentBib}

\bibitem [\protect \citeauthoryear {%
Duran~Vinent%
, Andreotti%
, Claudin%
\BCBL {}\ \BBA {} Winter%
}{%
Duran~Vinent%
\ \protect \BOthers {.}}{%
{\protect \APACyear {2019}}%
}]{%
duran2019unified}
\APACinsertmetastar {%
duran2019unified}%
\begin{APACrefauthors}%
Duran~Vinent, O.%
, Andreotti, B.%
, Claudin, P.%
\BCBL {}\ \BBA {} Winter, C.%
\end{APACrefauthors}%
\unskip\
\newblock
\APACrefYearMonthDay{2019}{}{}.
\newblock
{\BBOQ}\APACrefatitle {A unified model of ripples and dunes in water and
  planetary environments} {A unified model of ripples and dunes in water and
  planetary environments}.{\BBCQ}
\newblock
\APACjournalVolNumPages{Nature Geoscience}{12}{5}{345--350}.
\newblock
\begin{APACrefDOI} \doi{10.1038/s41561-019-0336-4} \end{APACrefDOI}
\PrintBackRefs{\CurrentBib}

\bibitem [\protect \citeauthoryear {%
Greeley%
, Marshall%
\BCBL {}\ \BBA {} Leach%
}{%
Greeley%
\ \protect \BOthers {.}}{%
{\protect \APACyear {1984}}%
}]{%
greeley1984microdunes}
\APACinsertmetastar {%
greeley1984microdunes}%
\begin{APACrefauthors}%
Greeley, R.%
, Marshall, J\BPBI R.%
\BCBL {}\ \BBA {} Leach, R\BPBI N.%
\end{APACrefauthors}%
\unskip\
\newblock
\APACrefYearMonthDay{1984}{}{}.
\newblock
{\BBOQ}\APACrefatitle {Microdunes and other aeolian bedforms on {V}enus: Wind
  tunnel simulations} {Microdunes and other aeolian bedforms on {V}enus: Wind
  tunnel simulations}.{\BBCQ}
\newblock
\APACjournalVolNumPages{Icarus}{60}{1}{152--160}.
\newblock
\begin{APACrefDOI} \doi{10.1016/0019-1035(84)90144-1} \end{APACrefDOI}
\PrintBackRefs{\CurrentBib}

\bibitem [\protect \citeauthoryear {%
Ho%
, Valance%
, Dupont%
\BCBL {}\ \BBA {} El~Moctar%
}{%
Ho%
\ \protect \BOthers {.}}{%
{\protect \APACyear {2014}}%
}]{%
ho2014aeolian}
\APACinsertmetastar {%
ho2014aeolian}%
\begin{APACrefauthors}%
Ho, T\BPBI D.%
, Valance, A.%
, Dupont, P.%
\BCBL {}\ \BBA {} El~Moctar, A\BPBI O.%
\end{APACrefauthors}%
\unskip\
\newblock
\APACrefYearMonthDay{2014}{}{}.
\newblock
{\BBOQ}\APACrefatitle {Aeolian sand transport: Length and height distributions
  of saltation trajectories} {Aeolian sand transport: Length and height
  distributions of saltation trajectories}.{\BBCQ}
\newblock
\APACjournalVolNumPages{Aeolian Research}{12}{}{65--74}.
\newblock
\begin{APACrefDOI} \doi{10.1016/j.aeolia.2013.11.004} \end{APACrefDOI}
\PrintBackRefs{\CurrentBib}

\bibitem [\protect \citeauthoryear {%
Lapotre%
\ \protect \BOthers {.}}{%
Lapotre%
\ \protect \BOthers {.}}{%
{\protect \APACyear {2016}}%
}]{%
lapotre2016large}
\APACinsertmetastar {%
lapotre2016large}%
\begin{APACrefauthors}%
Lapotre, M\BPBI G\BPBI A.%
, Ewing, R\BPBI C.%
, Lamb, M\BPBI P.%
, Fischer, W\BPBI W.%
, Grotzinger, J\BPBI P.%
, Rubin, D\BPBI M.%
\BDBL {}Yingst, R\BPBI A.%
\end{APACrefauthors}%
\unskip\
\newblock
\APACrefYearMonthDay{2016}{}{}.
\newblock
{\BBOQ}\APACrefatitle {Large wind ripples on {M}ars: A record of atmospheric
  evolution} {Large wind ripples on {M}ars: A record of atmospheric
  evolution}.{\BBCQ}
\newblock
\APACjournalVolNumPages{Science}{353}{6294}{55--58}.
\newblock
\begin{APACrefDOI} \doi{10.1126/science.aaf3206} \end{APACrefDOI}
\PrintBackRefs{\CurrentBib}

\bibitem [\protect \citeauthoryear {%
Lester%
}{%
Lester%
}{%
{\protect \APACyear {2026}}%
}]{%
lester2026data}
\APACinsertmetastar {%
lester2026data}%
\begin{APACrefauthors}%
Lester, C\BPBI W.%
\end{APACrefauthors}%
\unskip\
\newblock
\APACrefYearMonthDay{2026}{}{}.
\newblock
\APACrefbtitle {Data for ``{S}caling of impact ripples across planetary
  conditions''} {Data for ``{S}caling of impact ripples across planetary
  conditions''}\ [Data set].
\newblock
\APAChowpublished {Zenodo}.
\newblock
\begin{APACrefDOI} \doi{10.5281/zenodo.24285406} \end{APACrefDOI}
\PrintBackRefs{\CurrentBib}

\bibitem [\protect \citeauthoryear {%
Lester%
, Murray%
, Duran%
, Andreotti%
\BCBL {}\ \BBA {} Claudin%
}{%
Lester%
\ \protect \BOthers {.}}{%
{\protect \APACyear {2025}}%
}]{%
lester2025emergence}
\APACinsertmetastar {%
lester2025emergence}%
\begin{APACrefauthors}%
Lester, C\BPBI W.%
, Murray, A\BPBI B.%
, Duran, O.%
, Andreotti, B.%
\BCBL {}\ \BBA {} Claudin, P.%
\end{APACrefauthors}%
\unskip\
\newblock
\APACrefYearMonthDay{2025}{}{}.
\newblock
{\BBOQ}\APACrefatitle {Emergence of wind ripples controlled by mechanics of
  grain--bed impacts} {Emergence of wind ripples controlled by mechanics of
  grain--bed impacts}.{\BBCQ}
\newblock
\APACjournalVolNumPages{Nature Geoscience}{18}{4}{344--350}.
\newblock
\begin{APACrefDOI} \doi{10.1038/s41561-025-01672-w} \end{APACrefDOI}
\PrintBackRefs{\CurrentBib}

\bibitem [\protect \citeauthoryear {%
Miller%
, Marshall%
\BCBL {}\ \BBA {} Greeley%
}{%
Miller%
\ \protect \BOthers {.}}{%
{\protect \APACyear {1987}}%
}]{%
miller1987wind}
\APACinsertmetastar {%
miller1987wind}%
\begin{APACrefauthors}%
Miller, J\BPBI S.%
, Marshall, J\BPBI R.%
\BCBL {}\ \BBA {} Greeley, R.%
\end{APACrefauthors}%
\unskip\
\newblock
\APACrefYearMonthDay{1987}{}{}.
\newblock
{\BBOQ}\APACrefatitle {Wind ripples in low density atmospheres} {Wind ripples
  in low density atmospheres}.{\BBCQ}
\newblock
\APACjournalVolNumPages{NASA Technical Memorandum}{89810}{}{268--270}.
\PrintBackRefs{\CurrentBib}

\bibitem [\protect \citeauthoryear {%
P{\"a}htz%
\ \BBA {} Dur{\'a}n%
}{%
P{\"a}htz%
\ \BBA {} Dur{\'a}n%
}{%
{\protect \APACyear {2018}}%
}]{%
pahtz2018cessation}
\APACinsertmetastar {%
pahtz2018cessation}%
\begin{APACrefauthors}%
P{\"a}htz, T.%
\BCBT {}\ \BBA {} Dur{\'a}n, O.%
\end{APACrefauthors}%
\unskip\
\newblock
\APACrefYearMonthDay{2018}{}{}.
\newblock
{\BBOQ}\APACrefatitle {The cessation threshold of nonsuspended sediment
  transport across aeolian and fluvial environments} {The cessation threshold
  of nonsuspended sediment transport across aeolian and fluvial
  environments}.{\BBCQ}
\newblock
\APACjournalVolNumPages{Journal of Geophysical Research: Earth
  Surface}{123}{8}{1638--1666}.
\newblock
\begin{APACrefDOI} \doi{10.1029/2017jf004580} \end{APACrefDOI}
\PrintBackRefs{\CurrentBib}

\bibitem [\protect \citeauthoryear {%
P{\"a}htz%
\ \BBA {} Dur{\'a}n%
}{%
P{\"a}htz%
\ \BBA {} Dur{\'a}n%
}{%
{\protect \APACyear {2020}}%
}]{%
pahtz2020unification}
\APACinsertmetastar {%
pahtz2020unification}%
\begin{APACrefauthors}%
P{\"a}htz, T.%
\BCBT {}\ \BBA {} Dur{\'a}n, O.%
\end{APACrefauthors}%
\unskip\
\newblock
\APACrefYearMonthDay{2020}{}{}.
\newblock
{\BBOQ}\APACrefatitle {Unification of aeolian and fluvial sediment transport
  rate from granular physics} {Unification of aeolian and fluvial sediment
  transport rate from granular physics}.{\BBCQ}
\newblock
\APACjournalVolNumPages{Physical Review Letters}{124}{16}{168001}.
\newblock
\begin{APACrefDOI} \doi{10.1103/physrevlett.124.168001} \end{APACrefDOI}
\PrintBackRefs{\CurrentBib}

\bibitem [\protect \citeauthoryear {%
P{\"a}htz%
\ \BBA {} Dur{\'a}n%
}{%
P{\"a}htz%
\ \BBA {} Dur{\'a}n%
}{%
{\protect \APACyear {2023}}%
}]{%
pahtz2023scaling}
\APACinsertmetastar {%
pahtz2023scaling}%
\begin{APACrefauthors}%
P{\"a}htz, T.%
\BCBT {}\ \BBA {} Dur{\'a}n, O.%
\end{APACrefauthors}%
\unskip\
\newblock
\APACrefYearMonthDay{2023}{}{}.
\newblock
{\BBOQ}\APACrefatitle {Scaling laws for planetary sediment transport from
  {DEM-RANS} numerical simulations} {Scaling laws for planetary sediment
  transport from {DEM-RANS} numerical simulations}.{\BBCQ}
\newblock
\APACjournalVolNumPages{Journal of Fluid Mechanics}{963}{}{A20}.
\newblock
\begin{APACrefDOI} \doi{10.1017/jfm.2023.343} \end{APACrefDOI}
\PrintBackRefs{\CurrentBib}

\bibitem [\protect \citeauthoryear {%
Prigozhin%
}{%
Prigozhin%
}{%
{\protect \APACyear {1999}}%
}]{%
prigozhin1999nonlinear}
\APACinsertmetastar {%
prigozhin1999nonlinear}%
\begin{APACrefauthors}%
Prigozhin, L.%
\end{APACrefauthors}%
\unskip\
\newblock
\APACrefYearMonthDay{1999}{}{}.
\newblock
{\BBOQ}\APACrefatitle {Nonlinear dynamics of aeolian sand ripples} {Nonlinear
  dynamics of aeolian sand ripples}.{\BBCQ}
\newblock
\APACjournalVolNumPages{Physical Review E}{60}{1}{729--733}.
\newblock
\begin{APACrefDOI} \doi{10.1103/PhysRevE.60.729} \end{APACrefDOI}
\PrintBackRefs{\CurrentBib}

\bibitem [\protect \citeauthoryear {%
Rasmussen%
, Valance%
\BCBL {}\ \BBA {} Merrison%
}{%
Rasmussen%
\ \protect \BOthers {.}}{%
{\protect \APACyear {2015}}%
}]{%
rasmussen2015laboratory}
\APACinsertmetastar {%
rasmussen2015laboratory}%
\begin{APACrefauthors}%
Rasmussen, K\BPBI R.%
, Valance, A.%
\BCBL {}\ \BBA {} Merrison, J.%
\end{APACrefauthors}%
\unskip\
\newblock
\APACrefYearMonthDay{2015}{}{}.
\newblock
{\BBOQ}\APACrefatitle {Laboratory studies of aeolian sediment transport
  processes on planetary surfaces} {Laboratory studies of aeolian sediment
  transport processes on planetary surfaces}.{\BBCQ}
\newblock
\APACjournalVolNumPages{Geomorphology}{244}{}{74--94}.
\newblock
\begin{APACrefDOI} \doi{10.1016/j.geomorph.2015.03.041} \end{APACrefDOI}
\PrintBackRefs{\CurrentBib}

\bibitem [\protect \citeauthoryear {%
Schmerler%
, Katra%
, Kok%
, Tsoar%
\BCBL {}\ \BBA {} Yizhaq%
}{%
Schmerler%
\ \protect \BOthers {.}}{%
{\protect \APACyear {2016}}%
}]{%
schmerler2016experimental}
\APACinsertmetastar {%
schmerler2016experimental}%
\begin{APACrefauthors}%
Schmerler, E.%
, Katra, I.%
, Kok, J\BPBI F.%
, Tsoar, H.%
\BCBL {}\ \BBA {} Yizhaq, H.%
\end{APACrefauthors}%
\unskip\
\newblock
\APACrefYearMonthDay{2016}{}{}.
\newblock
{\BBOQ}\APACrefatitle {Experimental and numerical study of {S}harp’s shadow
  zone hypothesis on sand ripple wavelength} {Experimental and numerical study
  of {S}harp’s shadow zone hypothesis on sand ripple wavelength}.{\BBCQ}
\newblock
\APACjournalVolNumPages{Aeolian Research}{22}{}{37--46}.
\newblock
\begin{APACrefDOI} \doi{10.1016/j.aeolia.2016.05.006} \end{APACrefDOI}
\PrintBackRefs{\CurrentBib}

\bibitem [\protect \citeauthoryear {%
Ungar%
\ \BBA {} Haff%
}{%
Ungar%
\ \BBA {} Haff%
}{%
{\protect \APACyear {1987}}%
}]{%
ungar1987steady}
\APACinsertmetastar {%
ungar1987steady}%
\begin{APACrefauthors}%
Ungar, J\BPBI E.%
\BCBT {}\ \BBA {} Haff, P\BPBI K.%
\end{APACrefauthors}%
\unskip\
\newblock
\APACrefYearMonthDay{1987}{}{}.
\newblock
{\BBOQ}\APACrefatitle {Steady state saltation in air} {Steady state saltation
  in air}.{\BBCQ}
\newblock
\APACjournalVolNumPages{Sedimentology}{34}{2}{289--299}.
\newblock
\begin{APACrefDOI} \doi{10.1111/j.1365-3091.1987.tb00778.x} \end{APACrefDOI}
\PrintBackRefs{\CurrentBib}

\bibitem [\protect \citeauthoryear {%
Walker%
}{%
Walker%
}{%
{\protect \APACyear {1981}}%
}]{%
walker1981experimental}
\APACinsertmetastar {%
walker1981experimental}%
\begin{APACrefauthors}%
Walker, J\BPBI D.%
\end{APACrefauthors}%
\unskip\
\newblock
\APACrefYear{1981}.
\unskip\
\newblock
\APACrefbtitle {An experimental study of wind ripples} {An experimental study
  of wind ripples}\ \APACtypeAddressSchool {\BUPhD}{}{}.
\unskip\
\newblock
\APACaddressSchool {}{Massachusetts Institute of Technology}.
\PrintBackRefs{\CurrentBib}

\bibitem [\protect \citeauthoryear {%
Yizhaq%
\ \protect \BOthers {.}}{%
Yizhaq%
\ \protect \BOthers {.}}{%
{\protect \APACyear {2024}}%
}]{%
yizhaq2024coevolving}
\APACinsertmetastar {%
yizhaq2024coevolving}%
\begin{APACrefauthors}%
Yizhaq, H.%
, Tholen, K.%
, Saban, L.%
, Swet, N.%
, Lester, C.%
, Silvestro, S.%
\BDBL {}Katra, I.%
\end{APACrefauthors}%
\unskip\
\newblock
\APACrefYearMonthDay{2024}{}{}.
\newblock
{\BBOQ}\APACrefatitle {Coevolving aerodynamic and impact ripples on {E}arth}
  {Coevolving aerodynamic and impact ripples on {E}arth}.{\BBCQ}
\newblock
\APACjournalVolNumPages{Nature Geoscience}{17}{1}{66--72}.
\newblock
\begin{APACrefDOI} \doi{10.1038/s41561-023-01348-3} \end{APACrefDOI}
\PrintBackRefs{\CurrentBib}

\end{thebibliography}

\end{document}